\documentclass[sigconf,authorversion]{acmart2}

\usepackage{xcolor}

\usepackage{dirtytalk}
\usepackage{multirow}
\usepackage{float}

\AtBeginDocument{%
  \providecommand\BibTeX{{%
    \normalfont B\kern-.5em{\scshape i\kern-.25em b}\kern-.8em\TeX}}}

\setcopyright{rightsretained}
\acmConference[CHI '23 TIAM Workshop]{2023 CHI Conference on Human Factors in Computing Systems: Towards an Inclusive and Accessible Metaverse Workshop}{April 23, 2023}{Hamburg, Germany}
\acmBooktitle{2023 CHI Conference on Human Factors in Computing Systems: Towards an Inclusive and Accessible Metaverse Workshop (CHI '23 TIAM Workshop), April 23, 2023, Hamburg, Germany}

\usepackage{array}
\begin{document}

\newcommand{\PreserveBackslash}[1]{\let\temp=\\#1\let\\=\temp}
\newcolumntype{C}[1]{>{\PreserveBackslash\centering}p{#1}}
\newcolumntype{R}[1]{>{\PreserveBackslash\raggedleft}p{#1}}
\newcolumntype{L}[1]{>{\PreserveBackslash\raggedright}p{#1}}

\title[]{Towards a More Inclusive Metaverse via Designing Tools That Support Collaborative Virtual World Building by Users With and Without Disabilities}


\author{Ken Jen Lee and Edith Law}
\email{{kenjen.lee, edith.law}@uwaterloo.ca}
\affiliation{%
  \institution{University of Waterloo}
  \streetaddress{200 University Avenue West}
  \city{Waterloo}
  \state{Ontario}
  \country{Canada}
  \postcode{N2L 3G1}
}

\renewcommand{\shortauthors}{Lee and Law}

\begin{abstract}
Research has found social VR to bring various benefits to users with and without disabilities. 
Given the success of social VR applications that support user-created worlds, it is important to consider how we can empower users in building inclusive virtual worlds by investigating how tools for world building can be built to better support collaborations between users with and without disabilities.
As such, this position paper provides a brief discussion of existing research into important factors that should be considered during such collaborations, and possible future research directions.

\end{abstract}




\maketitle

\section{Introduction}



Social VR spaces serve many functions, they could provide safe social spaces to improve social skills \cite{maloney2020falling} while avoiding social anxiety \cite{rzeszewski2020virtual}, engage in activities meant for professional development or mental self-improvement \cite{maloney2020falling}, immersive cultural appreciation and educational activities (e.g., via interactions with users from other parts of the world) \cite{maloney2020falling}, a place to improve connectedness for those in long-distance relationships \cite{maloney2020falling}, and a comfortable environment that could act as a substitute for social interactions, especially during periods of physical isolation \cite{rzeszewski2020virtual}.
For users with limited mobility, social VR could make social events (e.g., concerts, group meditation) much more accessible \cite{maloney2020falling}.
Compared to more traditional social methods (e.g., instant messaging social networking applications), social VR applications provide immersive experiences that are closer to offline social communication and more diverse interactive modes through the support of richer non-verbal communication  \cite{wang2020social}.
Moreover, social activities that might be unique to social VR contexts have emerged, from mirror dwelling \cite{thibault2022mirror} to sleeping in social VR \cite{maloney2020falling}.

Beyond that, further possible benefits for people with disabilities could be inferred from research in the context of massively multiplayer online (MMO) games, which is currently more available than similar research in social VR contexts.
For example, when virtual worlds could be used anonymously, users can choose when, if at all, and how much they reveal to others about themselves, which could be a positive experience for anyone who might find it hard to socialize in real life \cite{Stendal_2010}.
It is also possible that for people with disabilities, social VR, just like MMO games, could improve their quality of life and sense of self-importance and allow them to participate in day-to-day activities \cite{stewart2010opportunities,kizelshteyn2008therapy}.
Friendships and communities formed in these virtual worlds could become an important emotional support network that can improve well-being and functioning ability \cite{wilson2006perceived, stewart2010opportunities}.
As Stewart et al. discussed in \cite{stewart2010opportunities}, in Second Life, communities of players with disabilities have also created specific virtual worlds for themselves, which serve as in-game locations for events, support and socializing. For example, there is an island named Brigadoon created for people with Asperger syndrome and autism, and a nightclub for people using wheelchairs named Wheelies.
Besides that, these communities also help refer newcomers to assistive hardware and software, and even create custom software for players with disabilities for whom the default game interfaces are unusable. 
Similar communities have also been founded in social VR applications. An example is Helping Hands, which is a VRChat community with the mission of giving an ``environment for both the hearing and deaf communities to socialize in a friendly and safe place'' \cite{helpinghands}.
Helping Hands has custom worlds in VRChat \cite{helpinghandsintro}, and offer sign language (SL) classes through these worlds. 
However, specific ``VR adaptation[s]'' are required for using SL in VR due to limitations associated with current consumer-grade VR tracking technologies \cite{helpinghands}.

A crucial feature needed to realize these discovered and potential benefits of social VR is the ability for users and communities to create their own customized worlds. The support for user-created worlds is arguable the reason why VRChat is one of the most used social VR applications \cite{thibault2022mirror}. Moreover, to make the world creation process easier, VRChat provides free tools like VRCSDK3 \cite{vrcsdk3} and Udon \cite{vrcudon}, which are used together with the Unity game engine.
Ideally, however, users should be able to build virtual worlds from within VR (which is already supported by existing social VR applications like Anyland \cite{anylandsteam}), with features that support smoother collaborations. 
Particularly, such tools should be able to support specific communities in collaboratively building worlds specific to their preferences, and for general worlds to be built from collaborations between users with and without disabilities so that their views and experiences are represented (similar to other contexts like policy and research \cite{johnson2003inclusive}). This could be an important step towards a decentralized way of building virtual worlds that are welcoming to all users, instead of resulting in the social isolation of users with disabilities.
As such, the \textbf{goal of this research} is to explore how collaborative virtual world building tools can be designed more inclusively to encourage effective collaborations between users with and without disabilities?


\section{Important Factors for Successful Collaborations Between People With and Without Disabilities}

Embregts et al. \cite{Embregts_2018} investigated what factors are considered important for successful collaborations between researchers with and without intellectual disabilities in the context of inclusive research.
They found that these factors include first building a mutual relationship of respect (e.g., personal boundaries), honesty enthusiasm and openness, effective communication (e.g., listening, giving and receiving feedback), achieving a collaboration that allows everyone involved to contribute (e.g., proper adjustments of working pace, clear goals and expectations), being aware of skills, development needs (e.g., an understanding of ``this is what I have to offer and this is what I find difficult'') and impact (e.g., understanding how intimidating new environments can be, availability of support networks).
Similarly, Brown et al. \cite{Brown_2020} researched important collaboration factors in the context of collaborating with students with disabilities as partners for testing a university's accessibility website.
Student collaborators with disabilities reported feeling low levels of confidence and self-doubt, and being used to not being consulted on accessibility services. As such, there was a learning curve for them to feel comfortable expressing their opinions as co-designers.

\section{Possible Research Directions}
Based on research discussed above, we discuss and speculate possible ways that collaborative VR world building tools/systems can be designed to be more inclusive.

\noindent\textbf{Building mutual relationships.} There are many aspects to building a mutually respectful and understanding relationship as collaborators. Within the context of collaborative VR world building, collaborations might happen between users who are not already acquainted with each other. In these cases, various support could be provided. 
For example, for new teams (who have no or little history collaborating together), the system can introduce an ice breaking segment before introducing the world building task. This segment could include various activities, like a note on the preferred respectful language (i.e., to respectfully refer to people with various disabilities \cite{Levy_2021}) if users with disabilities choose to disclose themselves, exchanging details on personal boundaries, collaboration preferences (e.g., if additional feedback is helpful), and expectations (e.g., an agreed upon working pace, the strengths and limitations of each user).
Other than that, prerecorded videos or text-based materials could be prepared and presented to users to encourage an open and respectful collaboration (e.g., VR applications like EchoVR and VRChat show such messages during loading screens).
This is important to provide users without disabilities the ``knowledge and practical skills that assist with social interactions with their disabled peers'' \cite{pivik2002using,roberts1999attitudes}
In terms of implementation, the ice breaking segment could take place in a customized ``ice breaking world'' separate from the ``world building world'' to better define the goal of each phase: getting to know each other in the former, and collaborating to build a virtual world in the latter.

\noindent\textbf{World building.}
Accessibility researchers have created various frameworks for designing inclusive technologies, e.g., ability-based design (ABD), universal design, universal usability, among others. 
Researchers could explore ways of incorporating such frameworks into the world building process in ways that are friendly and easy to use to users with little or no prior experience with accessibility design.
For example, one possible way of incorporating ABD, i.e., a design approach focused on users' abilities and how interfaces can be configured accordingly \cite{10.1145/1952383.1952384,nolte2022implementing}, is to define a standard where a user's abilities are automatically taken into consideration when UIs and in-world interactions are rendered to the user. For instance, if the default input to make the in-world menu appear is a downward swipe motion midair, the tool could make it easier to map alternative inputs for users of various abilities, e.g., pressing a button instead.
Users' preferred input devices could also be taken into consideration as well, i.e., input devices are not hard-coded during world building. 
Each user's abilities, in turn, can be configured using a one-time configuration tool (i.e., no need for reconfiguration every time users enter a new world).
Taking one step further, researchers could also explore how to build tools that allow world builders to incorporate interactions in their world that are not just inclusive, but also serve as physical therapy.
That said, while options to incorporate visual- and interaction-based configurations could make accessible virtual worlds easier to build, care must be taken to strike a balance.
Particularly, overly significant changes could potentially harm users' social experiences by removing common topics of discussion with other visitors of the virtual world, possibly resulting in unintended social isolation.

\noindent\textbf{Communication and workflow.} 
Similar to world building, future research could investigate how to borrow processes from existing frameworks to enhance the communication and workflow process. 
Let us take value sensitive design (VSD), which aims to guide ``insightful investigations into technological innovation in ways that foreground the well-being of human beings and the natural world'' \cite{friedman2019value}, as an example. 
How can collaborative world building tools empower stakeholders (especially those whose voices are usually suppressed, like users with disabilities) in communicating their values and better negotiate conflicts in values between users with and without disabilities?
This is important, given the possibility that users with disabilities might feel less confident with voicing their opinions in collaborative scenarios \cite{Brown_2020}.
In terms of workflow, VSD's methods, like the value sensitive action-reflection model, where designers are asked to improve their design iterations by taking into consideration values of other stakeholders through tools like envisioning cards and value scenarios, could be of inspiration. 
For example, the design process could be broken up into smaller chunks, and feedback can be given and received between each chunk.
As noted by Embregts et al. \cite{Embregts_2018}, including communication channels so that users could easily reach out to their support networks and others in their community is also an important consideration.
Moreover, new collaboration workflows could be investigated to better support users with varying working pace \cite{Embregts_2018}. 

While we propose a few possible ways in which the collaborative VR world building process could be made to be more inclusive and supportive of collaborations between users with and without disabilities, 
it is important to note that many of these ideas are mostly speculative. Ultimately, research methodologies akin to participatory design involving people (both experts and non-experts in building VR environments) with and without disabilities as fellow designers are crucial for impactful inclusive research.




\section{Statement of Contribution}
Given the various benefits of social VR for users with and without disabilities, it is reasonable to expect that its use will continue to expand. 
Towards increase the accessibility of the metaverse, we aim to contribute to the workshop by discussing ways in which all users are given equal access to tools that can help them build virtual worlds that are accessible. 
These tools could serve two purposes: i) to build community-specific virtual worlds, or ii) general use virtual worlds. 
Regardless, an important aspect of designing such world-building tools is to support effective collaborations between users with and without disabilities.
Given the importance of both biological and social dimensions of disabilities, we hope to complement discussions about devices and methodologies with questions about the future of empowering all users in collaboratively building accessible virtual worlds.

\bibliographystyle{ACM-Reference-Format}
\bibliography{main}


\begin{thebibliography}{22}


\ifx \showCODEN    \undefined \def \showCODEN     #1{\unskip}     \fi
\ifx \showDOI      \undefined \def \showDOI       #1{#1}\fi
\ifx \showISBNx    \undefined \def \showISBNx     #1{\unskip}     \fi
\ifx \showISBNxiii \undefined \def \showISBNxiii  #1{\unskip}     \fi
\ifx \showISSN     \undefined \def \showISSN      #1{\unskip}     \fi
\ifx \showLCCN     \undefined \def \showLCCN      #1{\unskip}     \fi
\ifx \shownote     \undefined \def \shownote      #1{#1}          \fi
\ifx \showarticletitle \undefined \def \showarticletitle #1{#1}   \fi
\ifx \showURL      \undefined \def \showURL       {\relax}        \fi
\providecommand\bibfield[2]{#2}
\providecommand\bibinfo[2]{#2}
\providecommand\natexlab[1]{#1}
\providecommand\showeprint[2][]{arXiv:#2}

\bibitem[Brown et~al\mbox{.}(2020)]%
        {Brown_2020}
\bibfield{author}{\bibinfo{person}{Kate Brown}, \bibinfo{person}{Alise~De Bie},
  \bibinfo{person}{Akshay Aggarwal}, \bibinfo{person}{Ryan Joslin},
  \bibinfo{person}{Sarah Williams-Habibi}, {and} \bibinfo{person}{Vipusaayini
  Sivanesanathan}.} \bibinfo{year}{2020}\natexlab{}.
\newblock \showarticletitle{Students with disabilities as partners: A case
  study on user testing an accessibility website}.
\newblock \bibinfo{journal}{\emph{International Journal for Students as
  Partners}} \bibinfo{volume}{4}, \bibinfo{number}{2} (\bibinfo{date}{oct}
  \bibinfo{year}{2020}), \bibinfo{pages}{97--109}.
\newblock
\urldef\tempurl%
\url{https://doi.org/10.15173/ijsap.v4i2.4051}
\showDOI{\tempurl}


\bibitem[Crow\_Se7en(2021)]%
        {helpinghandsintro}
\bibfield{author}{\bibinfo{person}{Crow\_Se7en}.}
  \bibinfo{year}{2021}\natexlab{}.
\newblock \bibinfo{booktitle}{\emph{Meet the Helping Hands | VRChat (American
  Sign Language)}}.
\newblock Crow\_Se7en.
\newblock
\urldef\tempurl%
\url{https://www.youtube.com/watch?v=QRiS2CQUpUg}
\showURL{%
Retrieved Feb 13, 2023 from \tempurl}


\bibitem[Embregts et~al\mbox{.}(2018)]%
        {Embregts_2018}
\bibfield{author}{\bibinfo{person}{Petri J. C.~M. Embregts},
  \bibinfo{person}{Elsbeth~F. Taminiau}, \bibinfo{person}{Luciënne Heerkens},
  \bibinfo{person}{Alice~P. Schippers}, {and} \bibinfo{person}{Geert van
  Hove}.} \bibinfo{year}{2018}\natexlab{}.
\newblock \showarticletitle{Collaboration in Inclusive Research: Competencies
  Considered Important for People With and Without Intellectual Disabilities}.
\newblock \bibinfo{journal}{\emph{Journal of Policy and Practice in
  Intellectual Disabilities}} \bibinfo{volume}{15}, \bibinfo{number}{3}
  (\bibinfo{date}{jul} \bibinfo{year}{2018}), \bibinfo{pages}{193--201}.
\newblock
\urldef\tempurl%
\url{https://doi.org/10.1111/jppi.12248}
\showDOI{\tempurl}


\bibitem[Friedman and Hendry(2019)]%
        {friedman2019value}
\bibfield{author}{\bibinfo{person}{Batya Friedman} {and}
  \bibinfo{person}{David~G Hendry}.} \bibinfo{year}{2019}\natexlab{}.
\newblock \bibinfo{booktitle}{\emph{Value sensitive design: Shaping technology
  with moral imagination}}.
\newblock \bibinfo{publisher}{Mit Press}.
\newblock


\bibitem[Hands(2020)]%
        {helpinghands}
\bibfield{author}{\bibinfo{person}{Helping Hands}.}
  \bibinfo{year}{2020}\natexlab{}.
\newblock \bibinfo{booktitle}{\emph{Helping Hands}}.
\newblock VRChat Legends Wiki.
\newblock
\urldef\tempurl%
\url{https://vrchat-legends.fandom.com/wiki/Helping_Hands}
\showURL{%
Retrieved Feb 13, 2023 from \tempurl}


\bibitem[Johnson and Walmsley(2003)]%
        {johnson2003inclusive}
\bibfield{author}{\bibinfo{person}{Kelley Johnson} {and} \bibinfo{person}{Jan
  Walmsley}.} \bibinfo{year}{2003}\natexlab{}.
\newblock \bibinfo{booktitle}{\emph{Inclusive research with people with
  learning disabilities: Past, present and futures}}.
\newblock \bibinfo{publisher}{Jessica Kingsley Publishers}.
\newblock


\bibitem[Kizelshteyn(2008)]%
        {kizelshteyn2008therapy}
\bibfield{author}{\bibinfo{person}{Mark Kizelshteyn}.}
  \bibinfo{year}{2008}\natexlab{}.
\newblock \showarticletitle{Therapy and the metaverse: Second
  Life{\textregistered} and the changing conditions of therapy for convalescent
  and chronically ill users}.
\newblock \bibinfo{journal}{\emph{Washington University Undergraduate Research
  Digest}} \bibinfo{volume}{4}, \bibinfo{number}{1} (\bibinfo{year}{2008}),
  \bibinfo{pages}{17--26}.
\newblock


\bibitem[Levy et~al\mbox{.}(2021)]%
        {Levy_2021}
\bibfield{author}{\bibinfo{person}{Lior Levy}, \bibinfo{person}{Qisheng Li},
  \bibinfo{person}{Ather Sharif}, {and} \bibinfo{person}{Katharina Reinecke}.}
  \bibinfo{year}{2021}\natexlab{}.
\newblock \showarticletitle{Respectful Language as Perceived by People with
  Disabilities}. In \bibinfo{booktitle}{\emph{The 23rd International {ACM}
  {SIGACCESS} Conference on Computers and Accessibility}}.
  \bibinfo{publisher}{{ACM}}.
\newblock
\urldef\tempurl%
\url{https://doi.org/10.1145/3441852.3476534}
\showDOI{\tempurl}


\bibitem[Maloney and Freeman(2020)]%
        {maloney2020falling}
\bibfield{author}{\bibinfo{person}{Divine Maloney} {and} \bibinfo{person}{Guo
  Freeman}.} \bibinfo{year}{2020}\natexlab{}.
\newblock \showarticletitle{Falling asleep together: What makes activities in
  social virtual reality meaningful to users}. In
  \bibinfo{booktitle}{\emph{Proceedings of the Annual Symposium on
  Computer-Human Interaction in Play}}. \bibinfo{pages}{510--521}.
\newblock


\bibitem[Manyland(2016)]%
        {anylandsteam}
\bibfield{author}{\bibinfo{person}{Anyland~+ Manyland}.}
  \bibinfo{year}{2016}\natexlab{}.
\newblock \bibinfo{booktitle}{\emph{Anyland Steam Page}}.
\newblock Anyland + Manyland.
\newblock
\urldef\tempurl%
\url{https://store.steampowered.com/app/505700/Anyland/}
\showURL{%
Retrieved Feb 13, 2023 from \tempurl}


\bibitem[Nolte et~al\mbox{.}(2022)]%
        {nolte2022implementing}
\bibfield{author}{\bibinfo{person}{Amelie Nolte}, \bibinfo{person}{Jacob
  Wobbrock}, \bibinfo{person}{Torben Volkmann}, {and} \bibinfo{person}{Nicole
  Jochems}.} \bibinfo{year}{2022}\natexlab{}.
\newblock \showarticletitle{Implementing Ability-Based Design: A Systematic
  Approach to Conceptual User Modeling}.
\newblock \bibinfo{journal}{\emph{ACM Transactions on Accessible Computing}}
  \bibinfo{volume}{15}, \bibinfo{number}{4} (\bibinfo{year}{2022}),
  \bibinfo{pages}{1--26}.
\newblock


\bibitem[Pivik et~al\mbox{.}(2002)]%
        {pivik2002using}
\bibfield{author}{\bibinfo{person}{Jayne Pivik}, \bibinfo{person}{Joan
  McComas}, \bibinfo{person}{Ian MaCfarlane}, {and} \bibinfo{person}{Marc
  Laflamme}.} \bibinfo{year}{2002}\natexlab{}.
\newblock \showarticletitle{Using virtual reality to teach disability
  awareness}.
\newblock \bibinfo{journal}{\emph{Journal of Educational Computing Research}}
  \bibinfo{volume}{26}, \bibinfo{number}{2} (\bibinfo{year}{2002}),
  \bibinfo{pages}{203--218}.
\newblock


\bibitem[Roberts and Smith(1999)]%
        {roberts1999attitudes}
\bibfield{author}{\bibinfo{person}{Clare~M Roberts} {and}
  \bibinfo{person}{Peta~R Smith}.} \bibinfo{year}{1999}\natexlab{}.
\newblock \showarticletitle{Attitudes and behaviour of children toward peers
  with disabilities}.
\newblock \bibinfo{journal}{\emph{International Journal of Disability,
  Development and Education}} \bibinfo{volume}{46}, \bibinfo{number}{1}
  (\bibinfo{year}{1999}), \bibinfo{pages}{35--50}.
\newblock


\bibitem[Rzeszewski and Evans(2020)]%
        {rzeszewski2020virtual}
\bibfield{author}{\bibinfo{person}{Micha{\l} Rzeszewski} {and}
  \bibinfo{person}{Leighton Evans}.} \bibinfo{year}{2020}\natexlab{}.
\newblock \showarticletitle{Virtual place during quarantine--a curious case of
  VRChat}.
\newblock \bibinfo{journal}{\emph{Rozw{\'o}j Regionalny i Polityka Regionalna}}
  \bibinfo{number}{51} (\bibinfo{year}{2020}), \bibinfo{pages}{57--75}.
\newblock


\bibitem[Stendal et~al\mbox{.}(2010)]%
        {Stendal_2010}
\bibfield{author}{\bibinfo{person}{Karen Stendal}, \bibinfo{person}{Susan
  Balandin}, {and} \bibinfo{person}{Judith Molka-Danielsen}.}
  \bibinfo{year}{2010}\natexlab{}.
\newblock \showarticletitle{Virtual worlds: A new opportunity for people with
  lifelong disability?}
\newblock \bibinfo{journal}{\emph{Journal of Intellectual
  {\&} Developmental Disability}} \bibinfo{volume}{36},
  \bibinfo{number}{1} (\bibinfo{date}{nov} \bibinfo{year}{2010}),
  \bibinfo{pages}{80--83}.
\newblock
\urldef\tempurl%
\url{https://doi.org/10.3109/13668250.2011.526597}
\showDOI{\tempurl}


\bibitem[Stewart et~al\mbox{.}(2010)]%
        {stewart2010opportunities}
\bibfield{author}{\bibinfo{person}{Stephanie Stewart}, \bibinfo{person}{Terri~S
  Hansen}, {and} \bibinfo{person}{Timothy~A Carey}.}
  \bibinfo{year}{2010}\natexlab{}.
\newblock \showarticletitle{Opportunities for people with disabilities in the
  virtual world of Second Life}.
\newblock \bibinfo{journal}{\emph{Rehabilitation Nursing}}
  \bibinfo{volume}{35}, \bibinfo{number}{6} (\bibinfo{year}{2010}),
  \bibinfo{pages}{254--259}.
\newblock


\bibitem[Thibault and Bujic(2022)]%
        {thibault2022mirror}
\bibfield{author}{\bibinfo{person}{Mattia Thibault} {and} \bibinfo{person}{Mila
  Bujic}.} \bibinfo{year}{2022}\natexlab{}.
\newblock \showarticletitle{" Mirror Dwellers": Social VR, Identity and
  Internet Culture}. In \bibinfo{booktitle}{\emph{DiGRA’22--Proceedings of
  the 2022 DiGRA International Conference: Bringing Worlds Together}}. DiGRA
  online library.
\newblock


\bibitem[VRChat(2023a)]%
        {vrcsdk3}
\bibfield{author}{\bibinfo{person}{VRChat}.} \bibinfo{year}{2023}\natexlab{a}.
\newblock \bibinfo{booktitle}{\emph{Choosing your SDK}}.
\newblock VRhat.
\newblock
\urldef\tempurl%
\url{https://docs.vrchat.com/docs/choosing-your-sdk}
\showURL{%
Retrieved Feb 11, 2023 from \tempurl}


\bibitem[VRChat(2023b)]%
        {vrcudon}
\bibfield{author}{\bibinfo{person}{VRChat}.} \bibinfo{year}{2023}\natexlab{b}.
\newblock \bibinfo{booktitle}{\emph{What is Udon?}}
\newblock VRhat.
\newblock
\urldef\tempurl%
\url{https://docs.vrchat.com/docs/what-is-udon}
\showURL{%
Retrieved Feb 11, 2023 from \tempurl}


\bibitem[Wang(2020)]%
        {wang2020social}
\bibfield{author}{\bibinfo{person}{Minhan Wang}.}
  \bibinfo{year}{2020}\natexlab{}.
\newblock \showarticletitle{Social VR: A New Form of Social Communication in
  the Future or a Beautiful Illusion?}. In \bibinfo{booktitle}{\emph{Journal of
  Physics: Conference Series}}, Vol.~\bibinfo{volume}{1518}. IOP Publishing,
  \bibinfo{pages}{012032}.
\newblock


\bibitem[Wilson et~al\mbox{.}(2006)]%
        {wilson2006perceived}
\bibfield{author}{\bibinfo{person}{Sylia Wilson}, \bibinfo{person}{Lindsay~A
  Washington}, \bibinfo{person}{Joyce~M Engel}, \bibinfo{person}{Marcia~A
  Ciol}, {and} \bibinfo{person}{Mark~P Jensen}.}
  \bibinfo{year}{2006}\natexlab{}.
\newblock \showarticletitle{Perceived social support, psychological adjustment,
  and functional ability in youths with physical disabilities.}
\newblock \bibinfo{journal}{\emph{Rehabilitation Psychology}}
  \bibinfo{volume}{51}, \bibinfo{number}{4} (\bibinfo{year}{2006}),
  \bibinfo{pages}{322}.
\newblock


\bibitem[Wobbrock et~al\mbox{.}(2011)]%
        {10.1145/1952383.1952384}
\bibfield{author}{\bibinfo{person}{Jacob~O. Wobbrock},
  \bibinfo{person}{Shaun~K. Kane}, \bibinfo{person}{Krzysztof~Z. Gajos},
  \bibinfo{person}{Susumu Harada}, {and} \bibinfo{person}{Jon Froehlich}.}
  \bibinfo{year}{2011}\natexlab{}.
\newblock \showarticletitle{Ability-Based Design: Concept, Principles and
  Examples}.
\newblock \bibinfo{journal}{\emph{ACM Trans. Access. Comput.}}
  \bibinfo{volume}{3}, \bibinfo{number}{3}, Article \bibinfo{articleno}{9}
  (\bibinfo{date}{apr} \bibinfo{year}{2011}), \bibinfo{numpages}{27}~pages.
\newblock
\showISSN{1936-7228}
\urldef\tempurl%
\url{https://doi.org/10.1145/1952383.1952384}
\showDOI{\tempurl}


\end{thebibliography}


\end{document}